\documentclass[aps, pra, amsmath, amssymb, reprint]{revtex4-1}

\usepackage[pdftex]{graphicx}
\usepackage[pdftex, hypertexnames=false]{hyperref}
\usepackage{bm, braket}

\begin{document}
\title{Dimensionality-induced BCS-BEC crossover in layered superconductors}
\author{Kyosuke Adachi}
\author{Ryusuke Ikeda}
\affiliation{Department of Physics, Kyoto University, Kyoto 606-8502, Japan}
\date{\today}
\begin{abstract}
Based on a simple model of a layered superconductor with strong attractive interaction, we find that the separation of the pair-condensation temperature from the pair-formation temperature becomes more remarkable as the interlayer hopping gets smaller. We propose from this result the BCS-BEC crossover induced by the change in dimensionality, for instance, due to insertion of additional insulating layers or application of uniaxial pressure. The emergence of a pseudogap in the electronic density of states, which supports the idea of the dimensionality-induced BCS-BEC crossover, is also verified.
\end{abstract}
\maketitle

\textit{Introduction}. The BCS-BEC crossover~\cite{Chen_Stajic_2005, Randeria_Taylor_2014} is an exciting phenomenon in Fermionic systems, which connects the condensation of weakly bound pairs described within the Bardeen-Cooper-Schrieffer (BCS) framework to the Bose-Einstein condensation (BEC) of strongly bound pairs. In the ultracold Fermi gases, the Feshbach resonance has made it possible to experimentally realize the BCS-BEC crossover by tuning the strength of the attractive interaction between the atoms. On the other hand, in most of the superconductors discovered so far, the attractive interaction between electrons is so weak that the electron-pair condensation is basically described within the BCS framework.

Recently, several experiments have suggested a surprising possibility that a strong attractive interaction may be present in the iron selenide (FeSe), one of the iron-based superconductors. In fact, in this material, the ratio of the superconducting-transition, or the pair-condensation, temperature $T_\mathrm{c}$ to the Fermi energy $E_\mathrm{F}$ is large especially in the electron band~\cite{Terashima_Kikugawa_2014, Kasahara_Watashige_2014}. In relation to this, the diamagnetic response has shown a strong superconducting-fluctuation effect~\cite{Kasahara_Yamashita_2016} and has been examined theoretically~\cite{Adachi_Ikeda_2017}. Further, the existence of the pseudogap has been suggested in the temperature region above $T_\mathrm{c}$ based on the NMR measurement~\cite{Shi_Arai_2017}.

In such a many-particle system with strong attractive interaction, it is expected that the BCS-BEC crossover can be experimentally induced by tuning the interaction strength. Material realization of the BCS-BEC crossover will open up an opportunity to elucidate unexplored physical properties in systems with strong attractive interaction: for example, transport properties and orbital magnetic-field effects, which are generally difficult to explore in trapped and neutral ultracold Fermi gases. In contrast to the ultracold Fermi gases, however, it is generally difficult to control the strength of the attractive interaction in superconductors. Therefore, another idea is required to induce the BCS-BEC crossover in such a superconductor with strong attractive interaction as FeSe.

In this study, we propose an idea that the BCS-BEC crossover may be caused by changing the dimensionality, for example, by inserting additional insulating layers or applying pressure uniaxially. By considering a model of a layered superconductor with strong attractive interaction, we calculate the pair-condensation temperature $T_\mathrm{c}$ and the pair-formation temperature $T^*$ based on the T-matrix approximation~\cite{Yanase_Yamada_1999, Maly_Janko_1999, Chen_Stajic_2005, Tsuchiya_Watanabe_2009}. We find that $T_\mathrm{c}$ and $T^*$ become more distant from each other as the dimensionality gets lower. In addition, on the basis of the same approximation, we show that the pseudogap appears in the electronic density of states when the interlayer hopping is small enough. These behaviors can be understood as the BCS-BEC crossover induced by the change in dimensionality.

\textit{Model}. We consider an attractive Hubbard model to describe many electrons moving on a simple tetragonal lattice:
\begin{eqnarray}
H &=& - t_{\parallel} \sum_{\langle i, j \rangle_{\parallel}, \sigma} c_{i \sigma}^\dag c_{j \sigma} - t_{\perp} \sum_{\langle i, j \rangle_{\perp}, \sigma} c_{i \sigma}^\dag c_{j \sigma} + \left( \mathrm{h. \, c.} \right) \nonumber \\
&&- U \sum_{i} c_{i \uparrow}^\dag c_{i \downarrow}^\dag c_{i \downarrow} c_{i \uparrow},
\label{Eq:Hamiltonian}
\end{eqnarray}
where $\langle i, j \rangle_{\parallel (\perp)}$ means intralayer (interlayer) nearest-neighbor bonds in the $a$-$b$ plane (along the $c$ axis), and correspondingly, $t_{\parallel} (> 0)$ and $t_{\perp} (> 0)$ are the intralayer- and the interlayer-hopping amplitudes, respectively. $U (> 0)$ is the strength of the attractive interaction, and $c_{i \sigma}^{(\dag)}$ represents the annihilation (creation) operator of an electron with spin $\sigma$ at the site $i$.

There are two kinds of independent dimensionless parameters in our Hamiltonian. One is the anisotropy ratio $r = t_{\perp} / t_{\parallel}$ ($\leq 1$), which controls the dimensionality. In the limit of $r \rightarrow 1$ ($r \rightarrow 0$), the system is purely three (two) dimensional. The other is the dimensionless attractive-interaction strength $u = U / t_{\parallel}$.

In the two-dimensional limit ($r \rightarrow 0$), the pair condensation at a finite temperature is expected to be replaced by the Berezinskii-Kosterlitz-Thouless (BKT) transition~\cite{Loktev_Quick_2001, Iskin_Melo_2009, Chubukov_eremin_2016, Bighin_Salasnich_2016, Bighin_Salasnich_2017, Mulkerin_2017, Matsumoto_Hanai_2018}. In this paper, we focus on a finite-$r$ regime ($r \gtrsim 0.05$) and do not discuss the BKT transition.

Though quasi-two-dimensional models~\cite{Iskin_Melo_2009} and anisotropic lattice models~\cite{Chen_Kosztin_1998, Chen_Kosztin_1999, Kornilovitch_2015} similar to Eq.~(\ref{Eq:Hamiltonian}) have been considered so far, the significant roles of the change in dimensionality has not been clarified. In addition, we stress that effects of the dimensionality change due to variation in the anisotropy ratio $r$ are different from finite-size effects caused by confinement in the $xy$ plane, which have been recently discussed in the context of ultracold Fermi gases~\cite{Toniolo_2017, Toniolo_2018}.

\begin{figure}[tbp]
\includegraphics[scale=0.7]{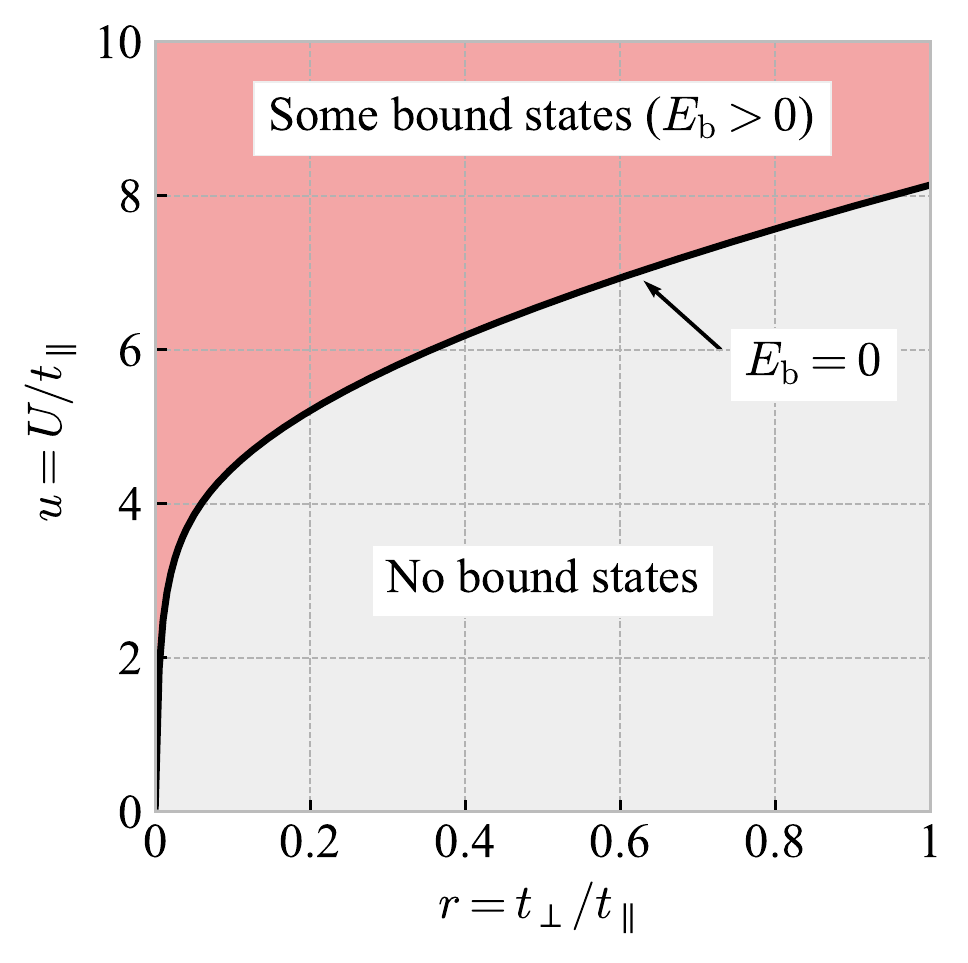}
\caption{The region where the two-particle bound state exists ($E_\mathrm{b} > 0$) in the $r$-$u$ plane (red area). The boundary satisfying $E_\mathrm{b} = 0$ (black line) and the region where no bound states exist (gray area) are also shown.}
\label{Fig:r-u}
\end{figure}

\textit{Formation of two-particle bound state}. Let us consider a two-particle system described by Eq.~(\ref{Eq:Hamiltonian}). If the attractive interaction is controlled in a many-particle system, the BCS-BEC crossover will take place when the interaction becomes strong enough to form a two-particle bound state~\cite{Randeria_Taylor_2014}. Thus, by solving the Schr{\" o}dinger equation of the corresponding two-particle system and calculating the threshold interaction strength for the bound-state formation, we can roughly estimate the characteristic interaction strength, at which the BCS-BEC crossover occurs in the many-particle system.

As shown in \cite{supplemental}, a bound state exists in the two-particle system described by Eq.~(\ref{Eq:Hamiltonian}) when the equation for the binding energy $E_\mathrm{b}$,
\begin{equation}
\frac{U}{M} \sum_{\bm{k}} \frac{1}{2 \epsilon_{\bm{k}} + W + E_\mathrm{b}} = 1,
\label{Eq:binding_energy}
\end{equation}
has a positive solution $E_\mathrm{b} > 0$. Here, we use the symbols $M = M_x M_y M_z$ as the number of lattice sites, $k_\alpha = 2 \pi n_\alpha / M_\alpha \, (\alpha = x, y, z)$ as the lattice momentum under the periodic boundary condition, $\epsilon_{\bm{k}} = -2 t_{\parallel} (\cos k_x + \cos k_y) - 2 t_{\perp} \cos k_z$ as the free-particle energy dispersion, and $W = 8 t_{\parallel} + 4 t_{\perp}$ as the band width. The binding energy $E_\mathrm{b}$ is measured from the bottom of the free-particle energy band.

In Fig.~\ref{Fig:r-u}, we show in the $r$-$u$ plane the red region where the two-particle bound state exists ($E_\mathrm{b} > 0$). The black line represents the boundary where the bound-state energy vanishes ($E_\mathrm{b} = 0$). If $u$ is changed under a fixed $r$, we obtain from Fig.~\ref{Fig:r-u} a certain value $u = u_0$, at which a bound state starts to appear (\textit{e}.\textit{g}., $u_0 = 6.58$ for $r = 0.5$). In the corresponding many-particle system, the BCS-BEC crossover is expected to occur when the interaction $u$ is tuned through $u_0$. On the other hand, if $r$ is changed under a fixed $u$, we find a certain value $r = r_0$, at which a bound state starts to appear (\textit{e}.\textit{g}., $r_0 = 0.356$ for $u = 6$). In the many-particle system, in the same way as the $u$-tuned case, we expect the BCS-BEC crossover to occur when the anisotropy ratio $r$ is changed, or the dimensionality is tuned, through $r_0$. This is our basic idea. In the following, we show that this scenario can be actually realized on the basis of the separation between the pair-formation temperature $T^*$ and the pair-condensation temperature $T_\mathrm{c}$ as well as the emergence of the pseudogap in the electronic density of states.

\begin{figure*}[tbp]
\includegraphics[scale=0.59]{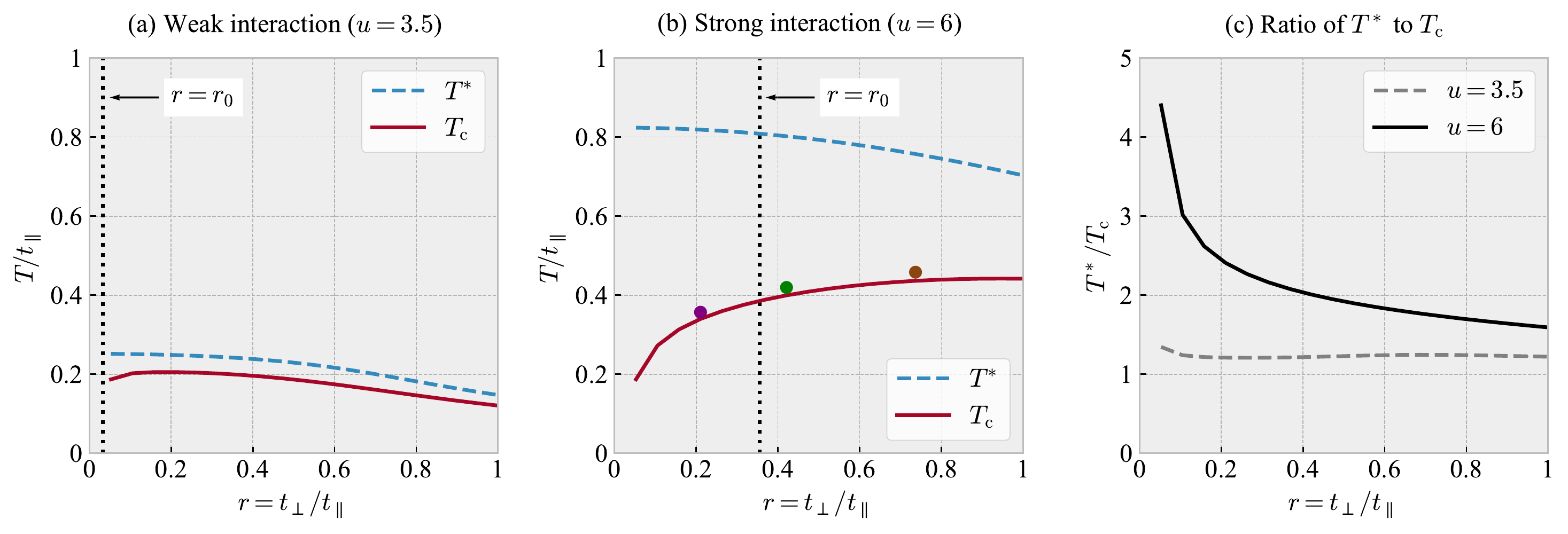}
\caption{The pair-formation temperature $T^*$ (blue dashed line) and the pair-condensation temperature $T_\mathrm{c}$ (red solid line) for systems with (a) weak interaction ($u = 3.5$) and (b) strong interaction ($u = 6$). At $r = r_0$ (black dotted line), the bound state starts to appear or vanish in the corresponding two-particle system. The colored points in (b) show $1.05 T_\mathrm{c}$ for each value of $r$, where the density of states are evaluated as shown in Fig.~\ref{Fig:DOS} below. In (c), the ratio of the pair-formation temperature $T^*$ to the pair-condensation temperature $T_\mathrm{c}$ is shown.}
\label{Fig:r-T}
\end{figure*}

\begin{figure}[tbp]
\includegraphics[scale=0.7]{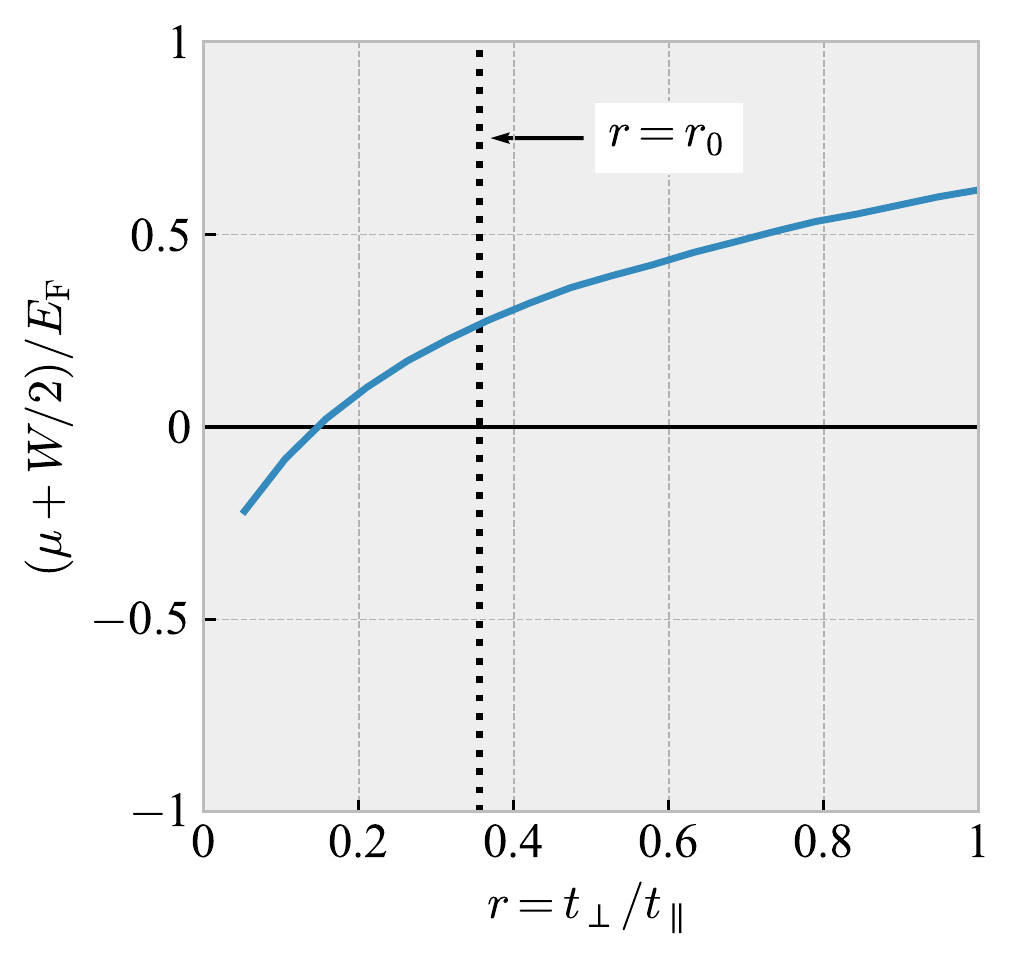}
\caption{The $r$ dependence of the chemical potential $\mu$ at the pair-condensation temperature $T_\mathrm{c}$ in the system with strong attractive interaction ($u = 6$). $W$ and $E_\mathrm{F}$ represent the band width and the free-particle Fermi energy measured from the band bottom, respectively. The black dotted line shows $r = r_0$ as in Fig.~\ref{Fig:r-T}(b).}
\label{Fig:r-mu}
\end{figure}

\textit{Separation between pair-formation and pair-condensation temperatures}. To show that the BCS-BEC crossover can occur through the change in dimensionality, we present the calculated results of the two characteristic temperatures, the pair-formation temperature $T^*$ and the pair-condensation temperature $T_\mathrm{c}$.

The pair formation is not a transition but a crossover phenomenon, and here we estimate $T^*$ based on the divergence of the uniform superconducting susceptibility $\chi_\mathrm{SC}$ within the mean-field approximation~\cite{Melo_Randeria_1993, Iskin_Melo_2009}. Introducing the free-particle Green's function $G^{(0)}_{\bm{k}} (\mathrm{i} \varepsilon_n) = (\mathrm{i} \varepsilon_n - \epsilon_{\bm{k}} + \mu)^{-1}$, the uniform superconducting susceptibility is written as $\chi_\mathrm{SC} = \chi_{\bm{0}}^{(0)} (0) [1 - U \chi_{\bm{0}}^{(0)} (0)]^{-1}$, where
\begin{equation}
\chi_{\bm{q}}^{(0)} (\mathrm{i} \omega_m) = \frac{T}{M} \sum_{\bm{k}, n} G_{\bm{k} + \bm{q}}^{(0)} (\mathrm{i} \varepsilon_n + \mathrm{i} \omega_m) G_{- \bm{k}}^{(0)} (- \mathrm{i} \varepsilon_n).
\end{equation}
Here, we use the symbols $T$ as the temperature and $\varepsilon_n = 2 \pi (n + 1 / 2) T$ ($\omega_m = 2 \pi m T$) as the Fermion (Boson) Matsubara frequency. We estimate $T^*$ by combining $U \chi_{\bm{0}}^{(0)} (0) = 1$ and the mean-field-level equation for the particle density $n$, $n = (2 / M) \sum_{\bm{k}} \{\exp[( \epsilon_{\bm{k}} - \mu) / T] + 1\}^{-1}$. In the strong-coupling limit ($u \rightarrow \infty$), we can easily show from the definitions that $T^* \propto |\mu| \propto U \propto E_\mathrm{b}$. Therefore, we can interpret $T^*$ as a temperature where the pair formation (or pair breaking) occurs even when the attractive interaction is strong.

The pair-condensation, or the superconducting-transition, temperature $T_\mathrm{c}$ is calculated within the T-matrix approximation~\cite{Yanase_Yamada_1999, Maly_Janko_1999, Chen_Stajic_2005, Tsuchiya_Watanabe_2009}. This approximation is qualitatively correct as long as the density $n$ is not so close to unity, and the chemical-potential shift is important. If $n$ is close to unity, and the filling is about one-half, the chemical-potential shift is not so important, and the interaction between the superconducting fluctuations is crucial. In this case, the self energy should be estimated within a more sophisticated method, \textit{e}.\textit{g}., the self-consistent T-matrix approximation~\cite{Haussmann_1994, Yanase_Yamada_1999, Chen_Stajic_2005}. In the following, therefore, we consider a relatively low-density system with $n = 0.2$.

Within the T-matrix approximation, as we explain in \cite{supplemental}, the pair-condensation temperature $T_\mathrm{c}$ is calculated by solving both the equation $U \chi_{\bm{0}}^{(0)} (0) = 1$ and the equation for the particle density $n$,
\begin{equation}
n = \frac{2 T}{M} \sum_{\bm{k}, n} G_{\bm{k}} (\mathrm{i} \varepsilon_n) \mathrm{e}^{+ \mathrm{i} \varepsilon_n 0}.
\label{Eq:number_eq}
\end{equation}
Here, the interacting-particle Green's function $G_{\bm{k}} (\mathrm{i} \varepsilon_n)$ is given as
\begin{equation}
G_{\bm{k}} (\mathrm{i} \varepsilon_n)^{-1} = G_{\bm{k}}^{(0)} (\mathrm{i} \varepsilon_n)^{-1}  - \Sigma_{\bm{k}} (\mathrm{i} \varepsilon_n),
\label{Eq:Dyson_eq}
\end{equation}
and the self energy $\Sigma_{\bm{k}} (\mathrm{i} \varepsilon_n)$ satisfies the following equation:
\begin{eqnarray}
\Sigma_{\bm{k}} (\mathrm{i} \varepsilon_n) &=& - \frac{T}{M} \sum_{\bm{q}, m} G_{\bm{q} - \bm{k}}^{(0)} (\mathrm{i} \omega_m - \mathrm{i} \varepsilon_n) \nonumber \\
&& \times \frac{U^2 \chi_{\bm{q}}^{(0)} (\mathrm{i} \omega_m)}{1 - U \chi_{\bm{q}}^{(0)} (\mathrm{i} \omega_m)} \mathrm{e}^{+ \mathrm{i} (\omega_m - \varepsilon_n) 0}.
\label{Eq:self_energy}
\end{eqnarray}
To consider the physical meaning of $T_\mathrm{c}$ estimated in the above formulas, let us consider the strong-coupling limit ($u \rightarrow \infty$) with $t_\parallel = t_\perp = t$. In this limit, we can obtain $T_\mathrm{c} \propto t^2 / U$, which corresponds to the BEC transition temperature of a non-interacting Bose system with a nearest-neighbor hopping $t_\mathrm{B} \propto t^2 / U$~\cite{Micnas_1990}. Therefore, $T_\mathrm{c}$ can be interpreted as the pair-condensation temperature even when the attractive interaction is strong.

The numerically calculated results of $T^*$ and $T_\mathrm{c}$ are summarized in Figs.~\ref{Fig:r-T}(a) and (b), which correspond to a system with weak interaction ($u = 3.5$) and a system with strong interaction ($u = 6$), respectively. The black dotted line shows the value of $r_0$, where the corresponding two-particle system begins to have a bound state.

In the case of $u = 3.5$ shown in Fig.~\ref{Fig:r-T}(a), where $r_0 \sim 0$, the separation between $T^*$ and $T_\mathrm{c}$ is small and does not change so much in a broad range of $r$. In fact, Fig.~\ref{Fig:r-T}(c) shows that the ratio of $T^*$ to $T_\mathrm{c}$ changes little for $u = 3.5$. This means that the pair formation and the pair condensation occur essentially at the same temperature as long as $r \gtrsim r_0$, and thus the BCS picture is applicable. 

In the case of $u = 6$ shown in Fig.~\ref{Fig:r-T}(b), the separation between $T^*$ and $T_\mathrm{c}$ becomes more remarkable as $r$ gets smaller through $r_0$. Actually, Fig.~\ref{Fig:r-T}(c) shows that for $u = 6$ the ratio of $T^*$ to $T_\mathrm{c}$ increases as $r$ decreases through $r_0$. The separation between $T^*$ and $T_\mathrm{c}$ indicates that the BCS-BEC crossover takes place along with the change in $r$, or the change in dimensionality.

We also present the $r$ dependence of the chemical potential $\mu$ at $T_\mathrm{c}$ for $u = 6$. As shown in Fig.~\ref{Fig:r-mu}, $\mu$ becomes lower than the bottom of the free-particle energy band ($\mu < - W / 2$) when $r$ is small enough. Since it is known that the chemical potential becomes lower than the band bottom through the BCS-BEC crossover~\cite{Nozieres_Schmitt-Rink_1985}, our result reinforces the scenario of the dimensionality-induced BCS-BEC crossover in the system with strong interaction.

\begin{figure}[tbp]
\includegraphics[scale=0.7]{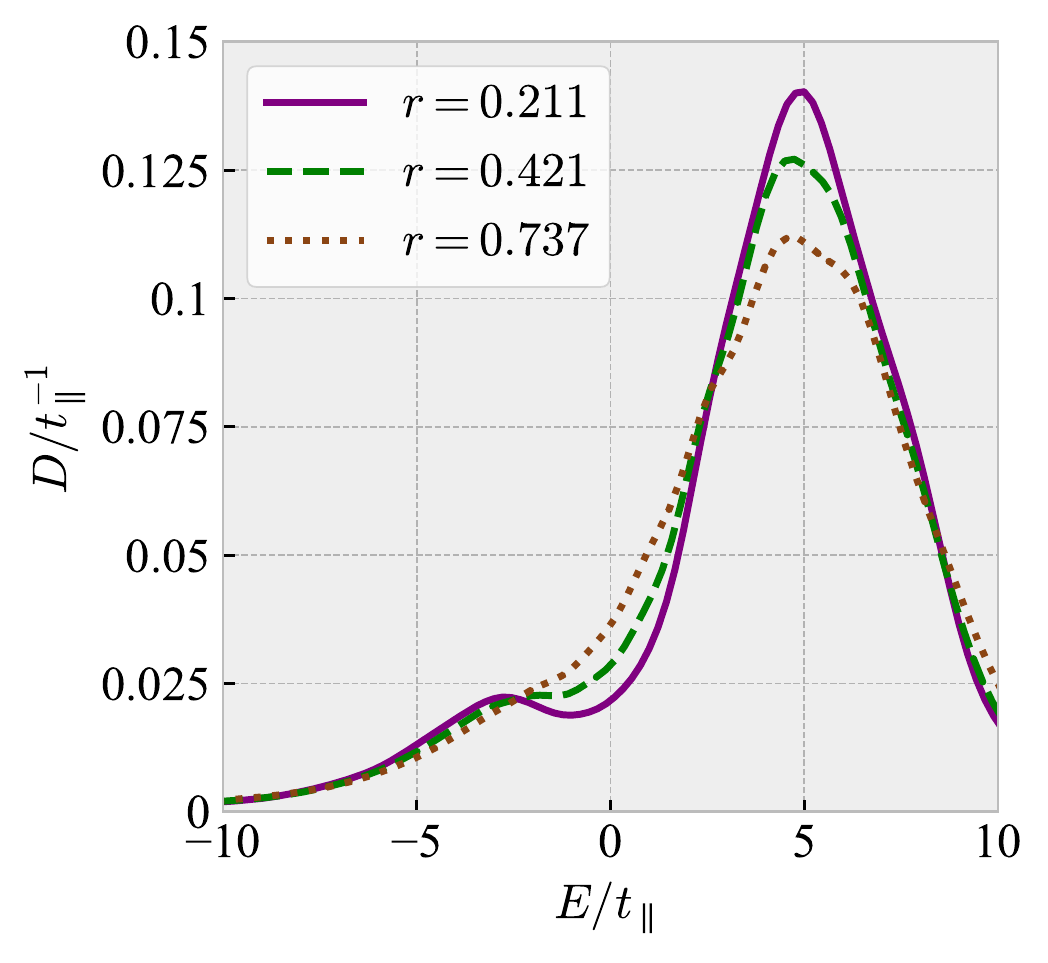}
\caption{The electronic density of states for the systems with strong attractive interaction ($u = 6$) and around the pair-condensation temperature ($T = 1.05 T_\mathrm{c}$). Each colored line corresponds to the colored point shown in Fig.~\ref{Fig:r-T}(b): $r = 0.211$ (purple solid line), $r = 0.421$ (green dashed line), and $r = 0.737$ (brown dotted line).}
\label{Fig:DOS}
\end{figure}

\textit{Pseudogap in electronic density of states}. To elucidate the effect of the dimensionality-induced BCS-BEC crossover on the one-particle excitation, we numerically calculate the electronic density of states $D (E)$ per spin per site. The calculation is based on the relation $D (E) = - \pi^{-1} \lim_{\gamma \rightarrow + 0} \mathrm{Im} G_{\bm{k}} (E + \mathrm{i} \gamma)$, where $G_{\bm{k}} (\mathrm{i} \varepsilon_n)$ is given in Eq.~(\ref{Eq:Dyson_eq}). The Pad{\' e} approximation is used for the analytic continuation from $G_{\bm{k}} (\mathrm{i} \varepsilon_n)$ to $G_{\bm{k}} (E + \mathrm{i} \gamma)$, and a finite energy width $\gamma = 0.1 W$ is introduced in the numerical calculation.

Figure~\ref{Fig:DOS} shows the obtained density of states $D (E)$ for the systems with $u = 6$. As shown with the colored points in Fig.~\ref{Fig:r-T}(b), we fix the temperature to $1.05 T_\mathrm{c}$ and change the anisotropy ratio $r$ . Figure~\ref{Fig:DOS} shows that the low-energy density of states becomes more depleted as $r$ gets smaller. The depletion of the density of states can be understood as the emergence of the pseudogap caused by the preformed-pair formation~\cite{Tsuchiya_Watanabe_2009}. Therefore, the behavior of the density of states is consistent with our picture of the dimensionality-induced BCS-BEC crossover. We note that the enhancement of the peak of $D(E)$ around $E / t_\parallel = 5$ in Fig.~\ref{Fig:DOS} basically originates from $r$ dependence of the non-interacting density of states in our model with $U = 0$ and thus is not always expected when the dimensionality-induced BCS-BEC crossover occurs.

According to studies on ultracold Fermi gases, theoretically as well as experimentally it is still controversial how a pseudogap is reflected in observables such as specific heat and magnetic susceptibility~\cite{Mueller_2017, Jensen_2018}.

\textit{Discussion}. We present the idea of the dimensionality-induced BCS-BEC crossover on the basis of a simple many-particle system described by Eq.~(\ref{Eq:Hamiltonian}). We find that the separation between $T^*$ and $T_\mathrm{c}$, as well as the depletion of the low-energy density of states, becomes prominent when the anisotropy ratio $r$ decreases through $r_0$. Here, $r_0$ is defined as a value of $r$, at which a bound state starts to appear in the corresponding two-particle system described by the same model [Eq.~(\ref{Eq:Hamiltonian})].

In more general classes of layered two-particle systems with $s$-wave attractive interaction, it is known that a two-particle bound state always exists in the two-dimensional limit, or the strong-anisotropy limit, regardless of the interaction strength~\cite{SchmittRink_Varma_1989}. Therefore, in such two-particle systems, the bound state is expected to appear when the anisotropy becomes sufficiently strong (as $r < r_0$ in our model). Accordingly, the idea of the dimensionality-induced BCS-BEC crossover can be naturally extended to the corresponding more general classes of layered many-particle system.

Regarding layered superconductors with strong attractive interaction such as FeSe, tuning the anisotropy may trigger the BCS-BEC crossover as discussed in this paper. As possible ways to control the anisotropy, we propose inserting additional insulating layers or applying uniaxial pressure/strain.

\textit{Acknowledgments}. One of the authors (K.~A.) is grateful to Y.~Ohashi, C.~A.~R.~S{\' a}~de~Melo, and J.~Ishizuka for fruitful discussions. The present research was supported by JSPS KAKENHI [Grants No.~16K05444 and No.~17J03883]. K.~A. also thanks JSPS for support from a Research Fellowship for Young Scientists.




%


\clearpage




\renewcommand{\thepage}{S\arabic{page}}
\setcounter{page}{1}
\renewcommand{\thefigure}{S\arabic{figure}}
\setcounter{figure}{0}
\renewcommand{\theequation}{S\arabic{equation}}
\setcounter{equation}{0}


\newcommand{\kbm}{\bm{k}}
\newcommand{\pbm}{\bm{p}}
\newcommand{\qbm}{\bm{q}}


\onecolumngrid

\begin{center}
{\large \bf Supplemental Material for ``Dimensionality-induced BCS-BEC crossover''}
\end{center}

\section{Equation for binding energy}

Let us consider the two-particle system described by the following Hamiltonian [Eq.~(1) in the main text]:
\begin{equation}
H = - t_{\parallel} \sum_{\langle i, j \rangle_{\parallel}, \sigma} \left( c_{i \sigma}^\dag c_{j \sigma} + c_{j \sigma}^\dag c_{i \sigma} \right) - t_{\perp} \sum_{\langle i, j \rangle_{\perp}, \sigma} \left( c_{i \sigma}^\dag c_{j \sigma} + c_{j \sigma}^\dag c_{i \sigma} \right) - U \sum_{i} c_{i \uparrow}^\dag c_{i \downarrow}^\dag c_{i \downarrow} c_{i \uparrow}.
\label{EqS:Hamiltonian}
\end{equation}
As explained in the main text, $\langle i, j \rangle_{\parallel (\perp)}$ means intralayer (interlayer) nearest-neighbor bonds in the $a$-$b$ plane (along the $c$ axis). In the same way, $t_{\parallel} (> 0)$ and $t_{\perp} (> 0)$ are the intralayer- and the interlayer-hopping amplitudes, respectively. $U (> 0)$ is the strength of the attractive interaction, and $c_{i \sigma}^{(\dag)}$ represents the annihilation (creation) operator of an electron with spin $\sigma$ at the site $i$.

To find the equation for the binding energy of a two-particle bound state, we start with the general two-particle state as a candidate for the eigenstate of Eq.~\eqref{EqS:Hamiltonian}:
\begin{equation}
\ket{\psi} = \sum_{\bm{k}, \bm{k'}} \sum_{\sigma, \sigma'} f_{\kbm \sigma, \kbm' \sigma'} \ket{\kbm \sigma, \kbm' \sigma'} = \sum_{\bm{k}, \bm{k'}} \sum_{\sigma, \sigma'} f_{\kbm \sigma, \kbm' \sigma'} \, c_{\kbm \sigma}^\dag c_{\kbm' \sigma'}^\dag \ket{0}.
\end{equation}
Here, $\ket{0}$ is the vacuum state, and the eigenfunction $f_{\kbm \sigma, \kbm' \sigma'}$ satisfies the antisymmetric relation
\begin{equation}
f_{\kbm' \sigma', \kbm \sigma} = -f_{\kbm \sigma, \kbm' \sigma'}.
\label{EqS:antisymmetric_relation}
\end{equation}
The Schr{\"o}dinger equation $H \ket{\psi} = E \ket{\psi}$, where $E$ is the eigenenergy, leads to the following equation:
\begin{equation}
\sum_{\kbm, \kbm'} \sum_{\sigma, \sigma'} (\epsilon_{\kbm} + \epsilon_{\kbm'}) f_{\kbm \sigma, \kbm' \sigma'} \, c_{\kbm \sigma}^\dag c_{\kbm' \sigma'}^\dag \ket{0} - \frac{2 U }{M} \sum_{\kbm, \kbm', \kbm''} f_{\kbm'' + \kbm + \kbm' \uparrow, -\kbm'' \downarrow} \, c_{\kbm \uparrow}^\dag c_{\kbm' \downarrow}^\dag \ket{0} = E \sum_{\kbm, \kbm'} \sum_{\sigma, \sigma'} f_{\kbm \sigma, \kbm' \sigma'} \, c_{\kbm \sigma}^\dag c_{\kbm' \sigma'}^\dag \ket{0}.
\label{EqS:Schrodinger_eq}
\end{equation}
Here, $\epsilon_{\kbm} = -2 t_\parallel (\cos k_x + \cos k_y) - 2 t_\perp \cos k_z$ is the free-particle energy dispersion.

For convenience, we split $f_{\kbm \uparrow, \kbm' \downarrow}$ into the symmetric part $f_{\kbm, \kbm'}^\mathrm{s}$ and the antisymmetric part $f_{\kbm, \kbm'}^\mathrm{a}$ as
\begin{equation}
f_{\kbm \uparrow, \kbm' \downarrow} = f_{\kbm, \kbm'}^\mathrm{s} + f_{\kbm, \kbm'}^\mathrm{a} = \frac{f_{\kbm \uparrow, \kbm' \downarrow} + f_{\kbm' \uparrow, \kbm \downarrow}}{2} + \frac{f_{\kbm \uparrow, \kbm' \downarrow} - f_{\kbm' \uparrow, \kbm \downarrow}}{2} = \frac{f_{\kbm \uparrow, \kbm' \downarrow} - f_{\kbm \downarrow, \kbm' \uparrow}}{2} + \frac{f_{\kbm \uparrow, \kbm' \downarrow} + f_{\kbm \downarrow, \kbm' \uparrow}}{2}.
\end{equation}
In the last equality, Eq.~\eqref{EqS:antisymmetric_relation} is used. Comparing the coefficients of $c_{\kbm \sigma}^\dag c_{\kbm' \sigma}^\dag \ket{0}$ ($\sigma = \uparrow, \downarrow$) in Eq.~\eqref{EqS:Schrodinger_eq} with one another, we obtain
\begin{equation}
(\epsilon_{\kbm} + \epsilon_{\kbm'}) f_{\kbm \sigma, \kbm' \sigma} = E f_{\kbm \sigma, \kbm' \sigma} \ \ (\sigma = \uparrow, \downarrow).
\label{EqS:Schrodinger_eq2}
\end{equation}
On the other hand, comparing the coefficients of $c_{\kbm \uparrow}^\dag c_{\kbm' \downarrow}^\dag \ket{0}$ in Eq.~\eqref{EqS:Schrodinger_eq} with one another, we obtain
\begin{equation}
(\epsilon_{\kbm} + \epsilon_{\kbm'}) f_{\kbm, \kbm'}^\mathrm{a} = E f_{\kbm, \kbm'}^\mathrm{a}
\label{EqS:Schrodinger_eq3}
\end{equation}
and
\begin{equation}
(\epsilon_{\kbm} + \epsilon_{\kbm'}) f_{\kbm, \kbm'}^\mathrm{s} - \frac{U}{M} \sum_{\kbm''} f_{\kbm'' + \kbm + \kbm', -\kbm''}^\mathrm{s} = E f_{\kbm, \kbm'}^\mathrm{s}.
\label{EqS:Schrodinger_eq4}
\end{equation}

First, we focus on $f_{\kbm \sigma, \kbm' \sigma}$ ($\sigma = \uparrow, \downarrow$) and $f_{\kbm, \kbm'}^\mathrm{a}$. Since they represent the eigenfunctions of the spin-triplet two-particle states, the singlet-channel attractive interaction $U$ does not work as seen in Eqs.~\eqref{EqS:Schrodinger_eq2} and \eqref{EqS:Schrodinger_eq3}, so that there are no bound states. Second, we focus on $f_{\kbm, \kbm'}^\mathrm{s}$. This eigenfunction corresponds to the spin-singlet two-particle state and is affected by the attractive interaction $U$ as seen in Eq.~\eqref{EqS:Schrodinger_eq4}, and thus a bound state may appear in $f_{\kbm, \kbm'}^\mathrm{s}$. Therefore, we discuss Eq.~\eqref{EqS:Schrodinger_eq4} in the following.

Let us assume that a bound state exists, so that the eigenenergy $E$ is below the free-particle ground-state energy $-W$, where $W = 8 t_\parallel + 4 t_\perp$ is the free-particle band width. The binding energy $E_\mathrm{b} (> 0)$ is defined as $E = -W - E_\mathrm{b}$. Defining $A_{\qbm} = M^{-1} \sum_{\kbm} f^\mathrm{s}_{\kbm + \qbm, -\kbm}$, we obtain from Eq.~\eqref{EqS:Schrodinger_eq4}
\begin{equation}
f^\mathrm{s}_{\kbm + \qbm, -\kbm} = \frac{U A_{\qbm}}{\epsilon_{\kbm + \qbm} + \epsilon_{-\kbm} + W + E_\mathrm{b}}.
\end{equation}
Summation over $\kbm$ in both sides of this equation leads to
\begin{equation}
\frac{U}{M} \sum_{\bm{k}} \frac{1}{\epsilon_{\kbm + \qbm} + \epsilon_{-\bm{k}} + W + E_\mathrm{b}} = 1.
\end{equation}
Since we are interested in the bound state with zero total momentum, we set $\qbm = \bm{0}$ and obtain the final expression for the binding energy $E_\mathrm{b}$ [Eq.~(2) in the main text]:
\begin{equation}
\frac{U}{M} \sum_{\bm{k}} \frac{1}{2 \epsilon_{\bm{k}} + W + E_\mathrm{b}} = 1.
\end{equation}

\section{T-matrix approximation}

For the sake of completeness, we explain the T-matrix approximation used to calculate the pair-condensation temperature $T_\mathrm{c}$. We introduce the free-particle Green's function
\begin{equation}
G^{(0)}_{\bm{k}} (\mathrm{i} \varepsilon_n) = \frac{1}{\mathrm{i} \varepsilon_n - \epsilon_{\bm{k}} + \mu},
\end{equation}
where $\varepsilon_n = 2 \pi (n + 1 / 2) T$ is the Fermion Matsubara frequency with temperature $T$, and $\mu$ is the chemical potential. The interacting-particle Green's function satisfies the following Dyson's equation:
\begin{equation}
G_{\kbm} (\mathrm{i} \varepsilon_n)^{-1} = G_{\kbm}^{(0)} (\mathrm{i} \varepsilon_n)^{-1} - \Sigma_{\kbm} (\mathrm{i} \varepsilon_n),
\label{EqS:Dyson_eq}
\end{equation}
where $\Sigma_{\kbm} (\mathrm{i} \varepsilon_n)$ is the self energy, which is estimated within the T-matrix approximation as explained in the following.

We define $\chi_{\bm{q}}^{(0)} (\mathrm{i} \omega_m)$ as
\begin{equation}
\chi_{\bm{q}}^{(0)} (\mathrm{i} \omega_m) = \frac{T}{M} \sum_{\bm{k}, n} G_{\bm{k} + \bm{q}}^{(0)} (\mathrm{i} \varepsilon_n + \mathrm{i} \omega_m) G_{- \bm{k}}^{(0)} (- \mathrm{i} \varepsilon_n),
\end{equation}
where $\omega_m = 2 \pi m T$ is the Boson Matsubara frequency. We also define the T matrix $T_{\qbm} (\mathrm{i} \omega_m)$ as
\begin{equation}
T_{\qbm} (\mathrm{i} \omega_m) = \frac{U}{1 - U \chi_{\bm{q}}^{(0)} (\mathrm{i} \omega_m)}.
\label{EqS:Tmatrix}
\end{equation}
Before discussing the T-matrix approximation, we consider the Hartree term:
\begin{equation}
\Sigma^\mathrm{(H)} = - U \frac{T}{M} \sum_{\kbm, n} G_{\kbm} (\mathrm{i} \varepsilon_n) \mathrm{e}^{+\mathrm{i} \varepsilon_n 0} = - \frac{U n}{2},
\end{equation}
where $n$ is the particle density. The contribution of this term to the self energy is constant if $U$ and $n$ are fixed. Thus, we take into account the Hartree term by properly choosing the origin of energy; in other words, we do not explicitly treat $\Sigma^\mathrm{(H)}$ in the expression of the self energy.

\begin{figure}[htbp]
\includegraphics[scale=0.4]{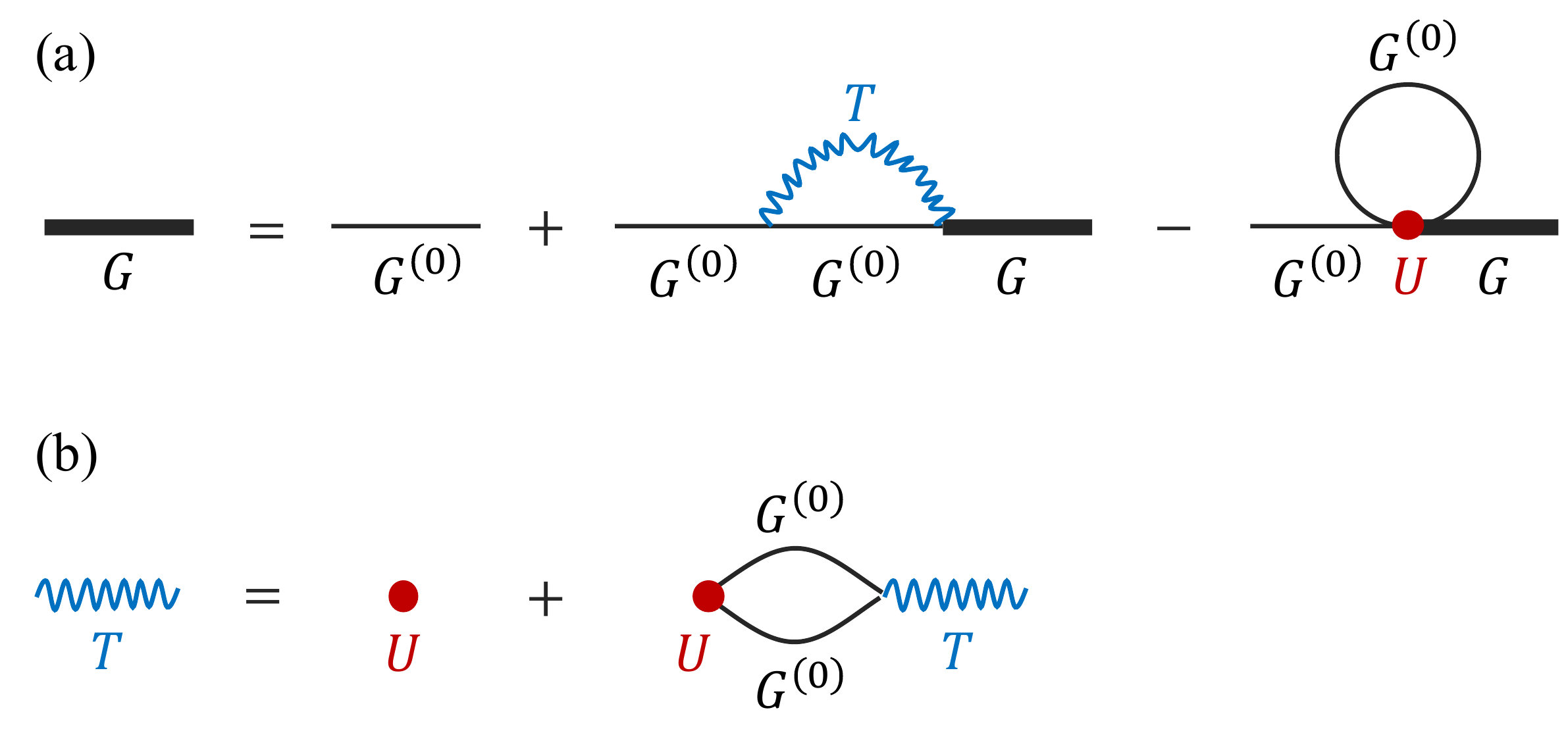}
\caption{The diagrammatic representation of Eqs.~\eqref{EqS:Dyson_eq}, \eqref{EqS:Tmatrix}, and \eqref{EqS:self_energy}. (a) The upper diagram shows the Dyson's equation, which gives the relation among the interacting-particle Green's function $G_{\kbm} (\mathrm{i} \varepsilon_n)$ (black bold line), the free-particle Green's function $G_{\kbm}^{(0)} (\mathrm{i} \varepsilon_n)$ (black thin line), the T matrix $T_{\qbm} (\mathrm{i} \omega_m)$ (blue wavy line), and the bare attractive interaction $U$ (red point). (b) The lower diagram expresses the recursive definition of the T matrix $T_{\qbm} (\mathrm{i} \omega_m)$.}
\label{FigS:diagram}
\end{figure}

Within the T-matrix approximation, the self energy is expressed as
\begin{equation}
\Sigma_{\kbm} (\mathrm{i} \varepsilon_n) = -\frac{T}{M} \sum_{\qbm, m} G_{\qbm - \kbm}^{(0)} (\mathrm{i} \omega_m - \mathrm{i} \varepsilon_n) T_{\qbm} (\mathrm{i} \omega_m) \mathrm{e}^{+\mathrm{i}(\omega_m - \varepsilon_n) 0} - \Sigma^{(1)},
\label{EqS:self_energy}
\end{equation}
where $\Sigma^{(1)} = - U (T / M) \sum_{\kbm, n} G_{\kbm}^{(0)} (\mathrm{i} \varepsilon_n) \mathrm{e}^{+\mathrm{i} \varepsilon_n 0}$ is the first-order perturbation term, which is implicitly taken into account in the Hartree term. Putting together the two terms in the right-hand side of Eq.~\eqref{EqS:self_energy}, we obtain the explicit representation of the self energy [Eq.~(6) in the main text]:
\begin{equation}
\Sigma_{\bm{k}} (\mathrm{i} \varepsilon_n) = - \frac{T}{M} \sum_{\bm{q}, m} G_{\bm{q} - \bm{k}}^{(0)} (\mathrm{i} \omega_m - \mathrm{i} \varepsilon_n) \frac{U^2 \chi_{\bm{q}}^{(0)} (\mathrm{i} \omega_m)}{1 - U \chi_{\bm{q}}^{(0)} (\mathrm{i} \omega_m)} \mathrm{e}^{+ \mathrm{i} (\omega_m - \varepsilon_n) 0}.
\end{equation}
Equations~\eqref{EqS:Dyson_eq}, \eqref{EqS:Tmatrix}, and \eqref{EqS:self_energy} are illustrated in Fig.~\ref{FigS:diagram} with the diagrammatic representation.

To consider the pair-condensation temperature $T_\mathrm{c}$ within the T-matrix approximation, we define the uniform superconducting susceptibility $\chi_\mathrm{SC}$ as
\begin{equation}
\chi_\mathrm{SC} = \frac{\chi_{\bm{0}}^{(0)} (0)}{1 - U \chi_{\bm{0}}^{(0)} (0)}.
\end{equation}
$T_\mathrm{c}$ is determined based on the divergence of $\chi_\mathrm{SC}$; in other words, we calculate $T_\mathrm{c}$ by solving the following equation:
\begin{equation}
U \chi_{\bm{0}}^{(0)} (0) = 1.
\label{EqS:Thouless_criterion}
\end{equation}
From Eq.~\eqref{EqS:Thouless_criterion}, we can obtain $T_\mathrm{c}$ as a function of the chemical potential $\mu$. Since we fix not the chemical potential $\mu$ but the number density $n$, we have to solve the following number equation [Eq.~(4) in the main text] together with Eq.~\eqref{EqS:Thouless_criterion} to determine the value of $\mu$:
\begin{equation}
n = \frac{T}{M} \sum_{\bm{k}, \sigma, n} G_{\bm{k}} (\mathrm{i} \varepsilon_n) \mathrm{e}^{+ \mathrm{i} \varepsilon_n 0} = \frac{2 T}{M} \sum_{\bm{k}, n} G_{\bm{k}} (\mathrm{i} \varepsilon_n) \mathrm{e}^{+ \mathrm{i} \varepsilon_n 0}.
\end{equation}







\end{document}